\documentclass[aps,prb,superscriptaddress,amsmath,amssymb,floatfix,reprint]{revtex4-2}
\usepackage{graphicx}
\usepackage{dcolumn}
\usepackage{bm}
\usepackage{physics}
\usepackage{color}
\usepackage[colorlinks=true, citecolor=blue, linkcolor=blue, urlcolor=blue]{hyperref}
\usepackage[T1]{fontenc}
\usepackage{tgtermes}

\begin{document}

\title{Coherent phonons in photoexcited semiconductors from first principles}
%
\author{Yuhui Qiu}
\author{Rong-Tian Pang}
\author{Meng-Xue Guan}
\author{Jin-Jian Zhou}
\email{jjzhou@bit.edu.cn}
\affiliation{Centre for Quantum Physics, Key Laboratory of Advanced Optoelectronic Quantum Architecture and Measurement (MOE), School of Physics, Beijing Institute of Technology, Beijing 100081, China.}
%

\begin{abstract}
	Coherent phonons are phase-locked lattice vibrations that provide a sensitive probe of nonequilibrium dynamics in materials. Existing first-principles equation-of-motion (EOM) approaches have successfully described coherent phonons in metals and semimetals, but their extension to semiconductors requires treating the long-lived electron--hole plasma created by above-gap excitation. 
	Here we extend the first-principles EOM framework to photoexcited semiconductors by incorporating the residual plasma force that shifts the lattice equilibrium together with the plasma-induced renormalization of the electronic structure.
	For monolayer MoTe$_2$, the calculated carrier-density-dependent Te displacement agrees with constrained density-functional perturbation theory
	only when the renormalized band structure is included.
	We then show that increasing the photoexcited carrier density enhances both the initial total force and the plasma-shifted equilibrium position of the
	$A_{1g}$ mode, producing larger coherent lattice oscillations.
	This density-controlled response appears in both the atomic displacement and the coherent-phonon-induced band renormalization, providing complementary
	signatures for ultrafast reflectivity and time- and angle-resolved photoemission spectroscopy.
\end{abstract}

\maketitle

\section{Introduction}

Coherent phonons are phase-locked collective lattice vibrations that are typically launched by ultrafast optical excitation.
Through their coupling to electronic, structural, magnetic, and optical degrees of freedom
, they provide a sensitive probe of nonequilibrium dynamics in quantum materials~\cite{Giannetti2016UltrafastOpticalSpectroscopy}. Since their early observation in pump-probe transient reflectivity experiments \cite{Thomsen1984CoherentPhononGeneration, Cheng1990ImpulsiveExcitationCoherent}, coherent phonons have been widely used to study ultrafast carrier and lattice relaxation \cite{Hase1998DynamicsCoherentPhonons, Zhang2024HighharmonicSpectroscopyProbes, Liu2022NanoscaleCoherentPhonon, Fu2025UltrafastCoherentPhonon}, structural phase transitions \cite{Guan2022OpticalControlMultistage, Horstmann2020CoherentControlSurface}, charge-ordered phases and charge density waves \cite{Gao2024ChargeOrderedPhases, Ratcliff2021CoherentPhononSpectroscopy, Zhong2024UnveilingVanHove}, superconductors \cite{Misochko2000DynamicsLowfrequencyPhonons}, metastable states \cite{Yang2021ExperimentalEvidenceMetastable, Forst2000PhaseChangeGe2Sb2Te5}, polarization dynamics \cite{Ma2023UltrafastGenerationDetection}, and thermal transport \cite{Luckyanova2012CoherentPhononHeat, Dettori2025CoherentPhononTransport}. 
Depending on the excitation conditions and material symmetry, coherent phonons can be generated through impulsive stimulated Raman scattering, ionic Raman scattering, infrared absorption, or the displacive excitation of coherent phonons (DECP) \cite{Zeiger1992TheoryDisplaciveExcitation}. 
In the above-gap regime relevant to semiconductors and many semimetals, DECP is often the dominant mechanism, in which photoexcited carriers abruptly modify the electronic force landscape, shift the lattice equilibrium, and launch coherent oscillations of symmetry-allowed modes \cite{Fukuda2024CoherentOpticalResponse, Fischer2025MicroscopicMechanismDisplacive}. Recent ultrafast experiments on semiconductors and layered materials, including transition-metal dichalcogenides, have further shown that coherent phonons can strongly modulate optical and electronic responses~\cite{Jensen2026CoherentPhononDrivenBand}, making them a powerful handle for probing and controlling nonequilibrium material properties~\cite{Sayers2023StrongCouplingCoherent, Fukuda2023PhotoinducedStructuralDynamics, GirottoErhardt2025UltrafastNonequilibriumEnhancement}.

First-principles approaches to coherent-phonon dynamics have developed rapidly alongside these experiments. 
Real-time time-dependent density functional theory directly describes coupled electron--lattice dynamics after photoexcitation and has provided important microscopic insight into coherent-phonon generation and dynamics~\cite{Shinohara2010CoherentPhononGeneration, Wang2025DirectionalPumpingCoherent}, but long-time relaxation and dephasing remain difficult to access within practical simulation times.
An alternative and complementary approach is the Heisenberg equation-of-motion (EOM) framework, which expresses coherent-phonon dynamics in terms of microscopic electron--phonon coupling and nonequilibrium carrier populations \cite{Caruso2023QuantumTheoryLightdriven, Stefanucci2023OutofequilibriumInitioTheory, Stefanucci2025FirstprinciplesEquationCoherent}. This framework unifies different coherent-phonon generation mechanisms within a common dynamical description and naturally incorporates damping. Recent first-principles studies have established this approach for metals and semimetals, describing DECP in BaFe$_2$As2$_2$~\cite{Lakehal2019MicroscopicDescriptionDisplacive}, quasiparticle band renormalization in antimony~\cite{Emeis2025CoherentPhononsQuasiparticle}, and coherent-phonon dephasing in bismuth and antimony~\cite{Pan2025InitioTheoryCoherent}. However, photoexcited semiconductors introduce a qualitatively different regime, in which above-gap excitation creates a long-lived electron--hole plasma that can persist throughout the coherent-phonon motion, so the lattice evolves in a nonequilibrium carrier background rather than around the original electronic state \cite{Fukuda2024CoherentOpticalResponse, Othonos1998ProbingUltrafastCarrier}. Incorporating this plasma into first-principles EOM simulations remains an important open step toward a predictive theory of coherent phonons in semiconductors~\cite{Caruso20252025RoadmapUltrafast}.

In this work, we extend the first-principles EOM approach for coherent phonons to the high-excitation regime of photoexcited semiconductors by 
incorporating the long-lived electron--hole plasma into both the lattice driving force and the plasma-renormalized electronic structure.
We apply the approach to monolayer MoTe$_2$, a prototypical two-dimensional semiconductor in which the out-of-plane $A_{1g}$ mode has been observed to strongly modulate the ultrafast optical response. We show that the electron--hole plasma substantially modifies the lattice response and provides a direct route to controlling the coherent phonon through the photoexcited carrier density. In particular, increasing the carrier density enhances the initial total force of the $A_{1g}$ mode, yielding larger-amplitude coherent oscillations and a larger plasma-shifted equilibrium. 
This density-controlled amplitude appears simultaneously in the time-dependent Te displacement and in the coherent-phonon-induced band renormalization, offering complementary signatures in ultrafast reflectivity and time- and angle-resolved photoemission spectroscopy.
Our results establish the carrier-density regime in which plasma-induced structural and electronic renormalization must be included in the description of coherent phonons and provide microscopic insight for interpreting coherent-phonon experiments in semiconductors.

\section{Theory and Methodology}

\subsection{Equation of motion for coherent phonons}

We first summarize the established first-principles equation of motion (EOM) for coherent phonons in the displacive excitation of coherent phonons (DECP) regime; the semiconductor-specific extension is introduced in the next subsection.
The time evolution of coherent phonons can be described by the atomic displacement $\Delta\tau_{\kappa p}(t)$, where $\kappa$ labels the atom and $p$ labels the unit cell. The displacement is expanded in normal coordinates $Q_{\bm{q}\nu}(t)$ as
\begin{equation}
	\label{tau}
	\Delta\tau_{\kappa p}(t)=\sum_{\bm{q}\nu}\left(\frac\hbar{2\omega_{\bm{q}\nu}M_\kappa}\right)^{\frac12}e^{i\bm{q}\cdot\bm{R}_p} \bm{e}_{\bm{q}\nu}^\kappa Q_{\bm{q}\nu}(t),
\end{equation}
where $\nu$ is the phonon mode index, $\omega_{\bm{q}\nu}$ is the phonon frequency, $M_\kappa$ is the atomic mass, $\bm{R}_p$ is the lattice vector, and $\bm{e}_{\bm{q}\nu}^\kappa$ is the phonon eigenvector.

For a spatially uniform DECP excitation, the macroscopic coherent displacement is carried by zone-center phonons \cite{Zeiger1992TheoryDisplaciveExcitation, Kuznetsov1994TheoryCoherentPhonon}, so we hereafter omit the phonon momentum index $\bm{q}$.
The normal-mode coordinate $Q_{\nu}(t)$ then obeys the Heisenberg EOM \cite{Emeis2025CoherentPhononsQuasiparticle, Caruso2023QuantumTheoryLightdriven}
\begin{subequations}
	\label{EOM}
	\begin{equation}
		\partial_{t}^{2}Q_{\nu}(t)+\gamma_{\nu}\partial_{t}Q_{\nu}(t)+\omega_{\nu}^{2}Q_{\nu}(t)=F_{\nu}(t),
	\end{equation}
	\begin{equation}
		Q_{\nu}(0)=0, \quad \partial_{t}Q_{\nu}(0)=0 .
	\end{equation}
\end{subequations}
Here, the scattering rate is $\gamma_{\nu} = 1 / \tau_{\nu}^{\mathrm{ph}}$, where $\tau_{\nu}^{\mathrm{ph}}$ is the phonon lifetime, and $F_{\nu}(t)$ is the force exciting coherent phonons.

In the established DECP formulation, this force originates from the photoexcited electrons and can be written as \cite{Emeis2025CoherentPhononsQuasiparticle, Caruso2023QuantumTheoryLightdriven}
\begin{equation}
	\label{EOM-force}
	F_{\nu}(t)=-2\omega_{\nu}\hbar^{-1}\sum_{n\bm{k}}g_{nn}^{\nu}(\bm{k},\bm{0})\left[f_{n\bm{k}}(t) - f_{n\bm{k}}^{(0)}\right],
\end{equation}
where $n$ is the band index, $\bm{k}$ is the electron momentum, $g_{nn}^\nu(\bm{k}, \bm{0})$ is the diagonal element of the electron--phonon coupling matrix at $\bm{q}=\Gamma$, $f_{n\bm{k}}(t)$ is the photoexcited electron distribution, and $f_{n\bm{k}}^{(0)}$ is the reference distribution against which the photoexcited population is measured.

For metals and semimetals, this reference distribution is naturally taken to be the equilibrium electronic occupation before optical excitation, denoted $f_{n\bm{k}}^\text{be}$. This choice underlies existing EOM descriptions of DECP in itinerant systems. Microscopic EOM work has described how photoinduced electronic feedback launches coherent phonons in metallic BaFe$_2$As$_2$ \cite{Lakehal2019MicroscopicDescriptionDisplacive}, while recent first-principles implementations compute the DECP force from ab initio electron--phonon matrix elements and reproduce the $A_{1g}$ coherent oscillation and associated quasiparticle band renormalization in semimetallic Sb \cite{Emeis2025CoherentPhononsQuasiparticle}. The same first-principles EOM framework has also been extended to coherent-phonon damping in semimetals, yielding dephasing rates consistent with measurements on Sb and Bi \cite{Pan2025InitioTheoryCoherent}.

In these metallic and semimetallic applications, the photoexcited distribution in Eq.~(\ref{EOM-force}) acts as a transient perturbation to $f_{n\bm{k}}^\text{be}$. Photoexcited carriers redistribute and effectively relax back to the pre-excitation distribution $f_{n\bm{k}}^\text{be}$ within a few picoseconds. During this interval, $F_\nu(t)$ shifts the instantaneous lattice equilibrium and launches the coherent phonon. As $f_{n\bm{k}}(t)-f_{n\bm{k}}^\text{be}$ decays, however, both the photoinduced force and the shifted equilibrium decay with it, and the damped oscillator ultimately returns to the original equilibrium structure.

\subsection{Coherent phonons in semiconductors}

In photoexcited semiconductors, an additional consideration is that the carrier distribution may not return to the pre-excitation state on the coherent-phonon timescale.
Before photoexcitation, the valence bands are fully occupied and the conduction bands are empty. Photoexcitation across the sizable band gap then generates electrons and holes that cannot recombine immediately; instead they accumulate near the band edges, forming an electron--hole plasma \cite{Fukuda2024CoherentOpticalResponse}. The plasma lifetime can exceed nanoseconds \cite{Othonos1998ProbingUltrafastCarrier}, far longer than the picosecond lifetime of the coherent phonon. On the timescale of the lattice dynamics, the plasma distribution is therefore quasi-static, and the lattice oscillates around a new equilibrium established by the plasma rather than around the pre-excitation structure.

This timescale separation changes the reference distribution used to define the force in Eq.~(\ref{EOM-force}). For the oscillatory coherent response, the force should vanish after the carriers have cooled into the long-lived electron--hole plasma. We therefore measure the transient part of the photoexcited distribution relative to the thermalized plasma distribution $f_{n\bm{k}}^\text{e-h}$, rather than relative to the equilibrium occupation before optical excitation, $f_{n\bm{k}}^\text{be}$. Here, $f_{n\bm{k}}^{\text{e-h}}$ denotes the electron occupation in the thermalized electron--hole plasma. It is constructed from Fermi--Dirac populations for conduction-band electrons and valence-band holes at the lattice temperature, with separate electron and hole quasi-Fermi levels chosen so that both carrier densities equal the imposed photoexcited density $n_c$. Carrier recombination is neglected on the coherent-phonon timescale.

For the lattice coordinate measured from the pre-excitation structure, however, the static difference between $f_{n\bm{k}}^\text{e-h}$ and $f_{n\bm{k}}^\text{be}$ must still be retained. The photoinduced force therefore separates the total force into a transient coherent component and a residual plasma component:
\begin{equation}
	F_\nu(t)=F_\nu^{\text{co}}(t)+F_\nu^0,
	\label{eq:f_decomp}
\end{equation}
where $F_\nu^{\text{co}}(t)$ is the driving force for coherent lattice vibrations and $F_\nu^{0}$ is the residual force that shifts the lattice equilibrium on the coherent-phonon timescale. These components are given by
\begin{equation}
	F_\nu^{\text{co}}(t)=-2\omega_{\nu}\hbar^{-1}\sum_{n\bm{k}}g_{nn}^{\nu}(\bm{k},\bm{0})\left[f_{n\bm{k}}(t)-f_{n\bm{k}}^\text{e-h}\right],
	\label{eq:f_co}
\end{equation}
and
\begin{equation}
	F_\nu^{0}=-2\omega_{\nu}\hbar^{-1}\sum_{n\bm{k}}g_{nn}^{\nu}(\bm{k},\bm{0})\left[f_{n\bm{k}}^\text{e-h}-f_{n\bm{k}}^\text{be}\right].
\end{equation}
As the photoexcited carrier distribution gradually relaxes toward $f_{n\bm{k}}^{\text{e-h}}$, the coherent force $F_\nu^{\text{co}}(t)$ decays to zero, while the residual plasma force $F_\nu^{0}$ remains. This residual force shifts the lattice equilibrium by a static offset relative to the pre-excitation structure, so we separate the normal coordinate into two components:
\begin{equation}
	Q_{\nu}(t) = Q_{\nu}^\text{co}(t)+Q_{\nu}^0 .
	\label{eq:cohQ}
\end{equation}
Here, $Q_\nu^\text{co}(t)$ is the oscillatory response about the plasma-shifted equilibrium, while $Q_\nu^0$ is the static displacement of that equilibrium.

Substituting Eq.~(\ref{eq:cohQ}) and Eq.~(\ref{eq:f_decomp}) into Eq.~(\ref{EOM}), with $Q_\nu^0$ time independent, gives the EOM for the coherent component:
\begin{equation}
	\partial_{t}^{2}Q_{\nu}^\text{co} + \gamma_{\nu} \partial_{t}Q_{\nu}^\text{co} + \omega_{\nu}^{2}Q_{\nu}^\text{co} = F_\nu^{\text{co}}(t)+(F_\nu^0-Q_{\nu}^0\omega_{\nu}^2).
	\label{EOM-semiconductor}
\end{equation}
The initial conditions for $Q_\nu^\text{co}(t)$ are $Q_\nu^\text{co}(0)=-Q_\nu^0$ and $\partial_t Q_\nu^\text{co}(0)=0$, since immediately after photoexcitation the ions are still at rest at the pre-excitation structure.

This decomposition also fixes the static limit of the semiconductor EOM. After the carrier distribution has relaxed to $f_{n\bm{k}}^\text{e-h}$, the coherent driving force vanishes, $F_\nu^\text{co}(t)\to 0$; after phonon damping, the oscillatory coordinate also vanishes, $Q_\nu^\text{co}(t)\to 0$. The remaining displacement is therefore the static plasma-induced offset $Q_\nu^0$. Imposing force balance at this plasma-shifted equilibrium in Eq.~(\ref{EOM-semiconductor}) gives $F_\nu^0-Q_\nu^0\omega_\nu^2=0$, and hence
\begin{equation}
	Q_{\nu}^0=\frac{F_\nu^{0}}{\omega_{\nu}^2}.
	\label{eq:Q0}
\end{equation}
The residual force $F_\nu^0$ therefore sets the displaced equilibrium, whereas the transient force $F_\nu^\text{co}(t)$ generates coherent oscillations about it. Together, $Q_\nu^0$ and $Q_\nu^\text{co}(t)$ determine the atomic displacement through Eq.~(\ref{tau}).

The remaining ingredient is the feedback of the electron--hole plasma on the electronic structure. The plasma changes the band energies and screening environment that determine the carrier occupations, so evaluating $f_{n\bm{k}}(t)$ and $f_{n\bm{k}}^{\text{e-h}}$ on the pre-excitation band structure would make the force inconsistent with the plasma state that generates it. We include this feedback through constrained density-functional perturbation theory (cDFPT), which treats the photoexcited semiconductor in the same constrained plasma state defined above \cite{Marini2021LatticeDynamicsPhotoexcited}. In this constrained state, cDFPT provides the static forces and relaxed structure associated with the long-lived plasma.

The relation between cDFPT and the present EOM is therefore complementary rather than redundant. cDFPT gives the plasma-shifted equilibrium but not the time-dependent coherent oscillation about it. The EOM uses the same plasma state to define $F_\nu^0$ and $F_\nu^\text{co}(t)$. In the static limit, after $F_\nu^\text{co}(t)$ has vanished and the oscillation has damped out, it must reduce to the cDFPT equilibrium condition $Q_\nu^0=F_\nu^0/\omega_\nu^2$. In practice, we use cDFPT-renormalized band structures to construct the carrier distributions entering both force components, and then use the direct cDFPT relaxed structure as an independent benchmark for the EOM prediction of $Q_\nu^0$.

\subsection{Computational details}

We apply the semiconductor EOM to monolayer MoTe$_2$. Spin-orbit coupling is included by employing fully relativistic norm-conserving pseudopotentials generated within the PBE functional by PSEUDO DOJO \cite{vanSetten2018PseudoDojoTrainingGrading}.First-principles calculations are performed using \textsc{Quantum ESPRESSO} with $12 \times 12 \times 1$ $\bm{k}$- and $\bm{q}$-point grids \cite{Giannozzi2009QUANTUMESPRESSOModular, Giannozzi2017AdvancedCapabilitiesMaterialsa}. To solve the equations of motion, the band energies and electron--phonon matrix elements are interpolated onto a denser $100 \times 100 \times 1$ grid using the \textsc{Perturbo} package \cite{Zhou2021PerturboSoftwarePackage}.
For polar electron-phonon interactions entering the carrier-cooling dynamics, the long-range contribution is treated using the two-dimensional Fr\"ohlich correction implemented in \textsc{Perturbo}.

On this dense grid, the time-dependent carrier distribution $f_{n\bm{k}}(t)$ is obtained by solving the Boltzmann transport equation with electron--phonon scattering using \textsc{Perturbo}. The carrier dynamics calculation includes eight valence bands and six conduction bands close to the Fermi energy. For the cDFPT calculations~\cite{Marini2021LatticeDynamicsPhotoexcited}, we impose the same thermalized electron--hole plasma state used as the EOM reference distribution $f_{n\bm{k}}^\text{e-h}$. The electron and hole occupations are constructed from separate Fermi--Dirac distributions at the lattice temperature of 295~K, with independent quasi-Fermi levels chosen so that the conduction-electron and valence-hole densities both equal $n_c$.
The lattice is treated as a damped harmonic oscillator with a carrier-density-independent mode frequency $\omega_\nu$ and damping rate $\gamma_\nu$. Carrier-induced and anharmonic frequency renormalization, as well as carrier-dependent phonon dephasing, are not included here; a fully first-principles treatment of these effects is beyond the scope of the present work. To highlight the decoherence process, we use a phenomenological phonon lifetime of 1~ps in our simulations, shorter than the experimentally measured damping time of  $\approx$6~ps for the $A_{1g}$ coherent phonon in monolayer MoTe$_2$ \cite{Sayers2023StrongCouplingCoherent}, so that the decay is clearly visible within the 2~ps simulation window.

Carrier recombination is neglected in our calculations. This fixed-density approximation is valid when the fractional carrier loss remains small over the simulation window. If recombination becomes appreciable on this timescale, the resulting decrease in carrier density would modify both the plasma-renormalized electronic structure and the coherent driving force and should be treated explicitly.

\section{Results}

\begin{figure}[!ht]
	\centering
	\includegraphics[width=\columnwidth]{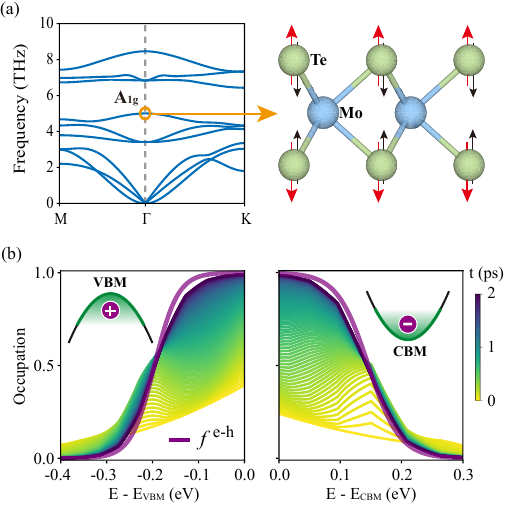}
	\caption{(a) Phonon spectrum of monolayer MoTe$_2$; the orange circle marks the zone-center $A_{1g}$ mode, whose out-of-plane displacement pattern is shown at right. (b) Relaxation of photoexcited carrier occupations from 0 to 2~ps. The left and right panels show hole occupations near the valence-band maximum and electron occupations near the conduction-band minimum, respectively. The initial state is a hot electron--hole distribution with density $n_c = 1.0 \times 10^{14}~\text{cm}^{-2}$ at $T = 2000~\text{K}$; the yellow-to-purple curves show the time evolution, and the thick purple curve denotes the thermalized electron--hole plasma reference $f_{n\bm{k}}^{\text{e-h}}$.}
	\label{a1}
\end{figure}

We apply the extended EOM framework to monolayer MoTe$_2$. 
Under spatially uniform above-gap excitation, the DECP force couples most directly to fully symmetric zone-center optical modes. 
We therefore focus on the out-of-plane $A_{1g}$ mode, highlighted in Fig.~\ref{a1}(a), consistent with recent coherent-phonon studies of monolayer MoTe$_2$ \cite{Fukuda2024CoherentOpticalResponse, Fukuda2023PhotoinducedStructuralDynamics, Sayers2023StrongCouplingCoherent}. 
In this mode, the two Te atoms move perpendicular to the monolayer plane with equal amplitude and opposite phase, as shown in Fig.~\ref{a1}(a). 
The computed frequency is approximately $5$~THz, corresponding to a period of about $200$~fs, in agreement with the measured value of $5.1$~THz \cite{Sayers2023StrongCouplingCoherent}.

The carrier dynamics are initialized after the rapid electron--electron thermalization stage. On a tens-of-femtoseconds timescale after photoexcitation, carrier-carrier scattering establishes hot quasi-equilibrium electron and hole populations, which we represent by Fermi--Dirac distributions \cite{Tong2021PreciseSimulationsCoupled, Othonos1998ProbingUltrafastCarrier},
\begin{equation}
	f^{\text{FD}}_{n\bm{k}}(\mu,T)=\{\exp[(E_{n\bm{k}}-\mu)/k_{B}T]+1\}^{-1}.
\end{equation}
For each imposed sheet density $n_c$, separate electron and hole quasi-Fermi levels are chosen so that the conduction-electron and valence-hole densities are equal to $n_c$. We then start from a hot distribution at $T = 2000~\text{K}$ and solve the Boltzmann transport equation with electron--phonon scattering. As shown in Fig.~\ref{a1}(b), the electron and hole occupations cool over the first 2~ps toward the thermalized electron--hole plasma reference $f_{n\bm{k}}^{\text{e-h}}$.
Note that we do not resolve the initial nonthermal carrier distribution before carrier-carrier scattering establishes a hot quasi-equilibrium population. In transition-metal dichalcogenides, this prethermal stage typically lasts only a few tens of femtoseconds, considerably shorter than the $\sim$200~fs period of the $A_{1g}$ phonon considered here.

\begin{figure}[!ht]
	\centering
	\includegraphics[width=\columnwidth]{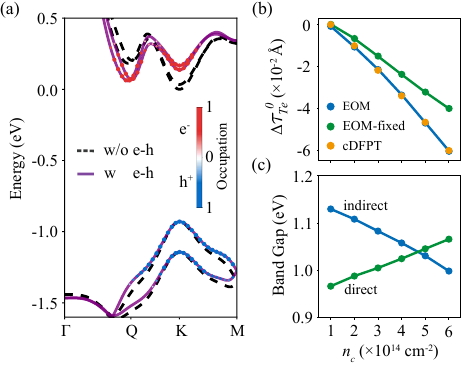}
	\caption{(a) Electronic band structures of monolayer MoTe$_2$ with and without the electron--hole plasma at $n_c = 6.0 \times 10^{14}~\text{cm}^{-2}$. The color scale marks carrier occupations in the constrained plasma state. (b) Equilibrium shift of the Te atoms as a function of $n_c$. The blue line is the EOM result including the plasma-induced electronic-structure modification, the green line is the fixed-band EOM result without this modification, and the orange circles are the cDFPT reference. (c) Direct band gap at $\bm{K}$ (green) and indirect band gap between $\bm{K}$ and $\bm{Q}$ (blue) as functions of $n_c$.}
	\label{band}
\end{figure}

The long-lived plasma also produces a sizable renormalization of the electronic structure. As shown in Fig.~\ref{band}(a), the bands computed in the constrained plasma state differ from the pre-excitation bands at $n_c = 6.0 \times 10^{14}~\text{cm}^{-2}$, with the strongest changes near the band edges where the photoexcited carriers accumulate. Since these same states determine $f_{n\bm{k}}^{\text{e-h}}$ and enter the electron--phonon force, the coherent-phonon calculation should use the plasma-renormalized bands rather than the pre-excitation band structure.

We first test the static limit of the plasma-renormalized force by comparing the predicted equilibrium displacement with cDFPT. In the extended EOM framework, the residual plasma force $F_{A_{1g}}^0$ produces an equilibrium shift $Q_{A_{1g}}^0=F_{A_{1g}}^0/\omega_{A_{1g}}^2$, which corresponds to a static Te displacement $\Delta\tau_{\text{Te}}^0$ through Eq.~(\ref{tau}). Figure~\ref{band}(b) shows that this displacement agrees with the cDFPT relaxed structure \cite{Marini2021LatticeDynamicsPhotoexcited} when the plasma-renormalized band structure is used.
If the force is evaluated on the pre-excitation bands instead, the displacement is underestimated, especially at high $n_c$.
The static comparison therefore validates the use of plasma-renormalized bands in the force construction before we examine the time-dependent coherent response.
Note that plasma-induced renormalization to the electron-phonon coupling is neglected in our calculations, it might also contribute to the remaining difference between the fixed-band EOM and cDFPT results in Fig.~\ref{band}(b), although including the band renormalization alone removes most of the discrepancy.

The same plasma-renormalized bands also reveal a density-dependent rearrangement of the band extrema. Figure~\ref{band}(c) shows that the direct gap at $\bm{K}$ increases with $n_c$, while the indirect gap between $\bm{K}$ and $\bm{Q}$ decreases. When $n_c$ exceeds approximately $5.0 \times 10^{14}~\text{cm}^{-2}$, the indirect gap becomes the smaller gap, showing that the conduction-band minimum moves from $\bm{K}$ to $\bm{Q}$.
This density-driven band-edge shift, together with the growth of the equilibrium lattice offset, identifies $n_c$ as the key control parameter for the coherent-phonon response, and also highlights that plasma-induced structural and electronic renormalization must be included to accurately describe the coherent-phonon dynamics at the high-excitation regime.

\begin{figure}[!ht]
	\centering
	\includegraphics[width=\columnwidth]{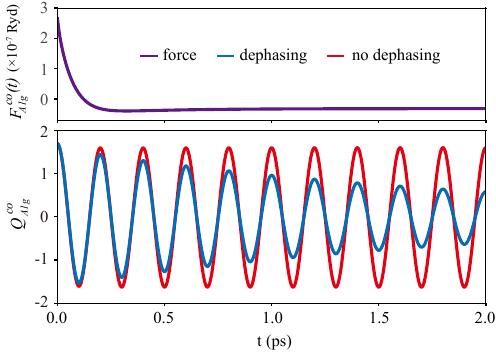}
	\caption{Time evolution of the $A_{1g}$ coherent phonon for photoexcited carriers with $n_c = 4.0 \times 10^{14}~\text{cm}^{-2}$ and initial carrier temperature $T = 1000~\text{K}$. The upper panel shows the transient coherent force $F_{A_{1g}}^{\text{co}}(t)$, which decays as the carrier distribution approaches $f_{n\bm{k}}^{\text{e-h}}$. The lower panel shows the coherent normal coordinate $Q_{A_{1g}}^{\text{co}}(t)$ with dephasing ($\gamma_{A_{1g}} > 0$, blue) and without dephasing ($\gamma_{A_{1g}} = 0$, red).}
	\label{force}
\end{figure}

We then use the plasma-shifted equilibrium as the reference point for the time-dependent coherent motion. Figure~\ref{force} shows the transient coherent force $F_{A_{1g}}^{\text{co}}(t)$ and the resulting coordinate $Q_{A_{1g}}^{\text{co}}(t)$.
According to Eq.~(\ref{eq:f_co}), the coherent force $F_\nu^{\mathrm{co}}(t)$ measures the deviation of the instantaneous carrier distribution $f_{n\bm{k}}(t)$ from the thermalized plasma distribution $f_{n\bm{k}}^{\mathrm{e-h}}$, weighted by the electron-phonon coupling $g_{nn}^{\nu}(\bm{k},\bm{0})$. As the hot carriers cool toward $f_{n\bm{k}}^{\mathrm{e-h}}$, this deviation diminishes and $F_\nu^{\mathrm{co}}(t)$ consequently approaches zero.
The small negative value remaining at 2~ps reflects incomplete carrier cooling within the simulation window [see Fig.~\ref{a1}(b)] and vanishes as the carrier distribution further relaxes toward $f_{n\bm{k}}^{\text{e-h}}$.
The lattice then oscillates about the plasma-shifted equilibrium at the $A_{1g}$ frequency. In the absence of dephasing the oscillation amplitude is conserved, whereas the phenomenological phonon lifetime damps the response. This example separates the static offset $Q_{A_{1g}}^0$ from the initial driving force $F_{A_{1g}}(0)$, both of which vary with carrier density and shape the coherent response shown below.

We next examine how these quantities scale with the photoexcited carrier density.
Experimentally, coherent phonons can be tuned by changing the photon energy and intensity of the incident laser \cite{Garl2008BirthDecayCoherent, Fischer2025MicroscopicMechanismDisplacive, Ishioka2024SuppressionShearIonic, Li2024AnisotropicPhononDynamics, Fukuda2023PhotoinducedStructuralDynamics, Ishioka2008CoherentA1gEg, Ishioka2006TemperatureDependenceCoherent, Ishioka2023SelectiveCouplingCoherent, Yamamoto1994CoherentOscillationZonefolded}. Higher photon energies and stronger intensities generate photoexcited carriers at higher densities \cite{Fischer2025MicroscopicMechanismDisplacive, Emeis2025CoherentPhononsQuasiparticle}, so $n_c$ provides the natural control parameter for the density sweep of the coherent-phonon response.

\begin{figure}[!ht]
	\centering
	\includegraphics[width=\columnwidth]{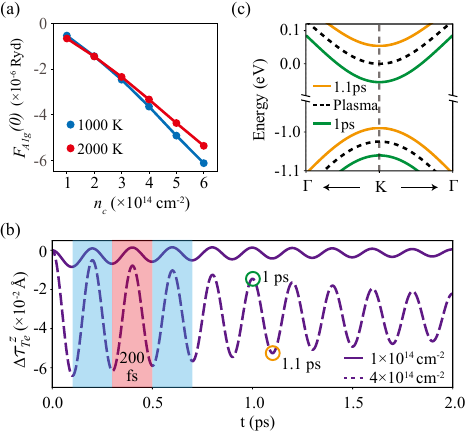}
	\caption{(a) Initial total force $F_{A_{1g}}(0)$ of the $A_{1g}$ mode as a function of photoexcited carrier density $n_c$ at initial carrier temperatures $T = 1000~\text{K}$ (blue) and $T = 2000~\text{K}$ (red). The force keeps the same sign over the plotted density range and grows in magnitude with $n_c$. (b) Time evolution of the Te displacement $\Delta\tau_\text{Te}^\text{z}(t)$ at $T = 1000~\text{K}$ for $n_c = 1.0 \times 10^{14}~\text{cm}^{-2}$ (solid) and $n_c = 4.0 \times 10^{14}~\text{cm}^{-2}$ (dashed). Both trajectories oscillate with the same period and phase, while the higher-density case has a larger amplitude and a larger equilibrium offset. Green and orange circles mark the higher-density displacement at 1.0~ps and 1.1~ps. (c) Coherent-phonon-induced band renormalization near $\bm{K}$ at 1.0~ps (green) and 1.1~ps (orange), compared with the plasma-reference bands (black dashed lines).}
	\label{cont}
\end{figure}

Within the EOM, varying $n_c$ changes the microscopic carrier occupations and the plasma-renormalized band structure on which the force is evaluated. These changes affect two distinct quantities: the residual plasma force $F^0$, which sets the shifted equilibrium [see Eq.~(\ref{eq:Q0})], and the initial total force $F(t=0)$, which launches the coherent oscillation. A density sweep therefore separates how the long-lived electron--hole plasma controls the static lattice offset and the initial oscillatory drive.

Figure~\ref{cont}(a) shows how the initial total force evolves with carrier density. For both initial carrier temperatures, $T=1000~\text{K}$ and $T=2000~\text{K}$, $F_{A_{1g}}(0)$ keeps the same sign while its magnitude grows with $n_c$. The unchanged sign means that the lattice is launched in the same direction over the density range considered, whereas the increasing magnitude gives a stronger coherent drive. Combined with the larger plasma-shifted displacement in Fig.~\ref{band}(b), this trend shows that increasing $n_c$ enhances the coherent response without producing a phase reversal. At high density, the same carrier range overlaps with the $\bm{K}$-to-$\bm{Q}$ band-edge rearrangement in Fig.~\ref{band}(c), which can suppress radiative recombination and further lengthen the electron--hole plasma lifetime \cite{Othonos1998ProbingUltrafastCarrier}.

The density dependence of these two force components is directly reflected in the atomic displacement. Although the normal coordinate $Q_{\nu}(t)$ is not itself an observable, it gives the real-space displacement through Eq.~(\ref{tau}). In the semiconductor case, this displacement includes both the plasma-induced equilibrium offset and the oscillatory coherent motion [see Eq.~(\ref{eq:cohQ})]. Figure~\ref{cont}(b) compares $\Delta\tau_\text{Te}^\text{z}(t)$ for $n_c = 1.0 \times 10^{14}~\text{cm}^{-2}$ and $n_c = 4.0 \times 10^{14}~\text{cm}^{-2}$. The two trajectories have the same period, $\omega_{A_{1g}}^{-1}\approx 200$~fs, and remain in phase, but the higher-density trajectory shows both a larger equilibrium offset and a larger oscillation amplitude. The offset follows the increase of $Q_{A_{1g}}^0$, while the larger amplitude reflects the stronger initial drive. Such density-enhanced displacements can be probed by time-resolved reflectivity, for which $\Delta I \sim \Delta\tau$ at small displacement \cite{Fukuda2024CoherentOpticalResponse}.

The coherent motion also induces a time-dependent band renormalization through the electron self-energy \cite{Lafuente-Bartolome2022UnifiedApproachPolarons, Emeis2025CoherentPhononsQuasiparticle, Wang2025DirectionalPumpingCoherent}:
\begin{equation}
	\label{selfenergy}
	\Sigma_{n\bm{k}}(t)=\sum_{\bm{q}\nu}g_{nn}^\nu(\bm{k},\bm{q})Q^\text{co}_{\bm{q}\nu}(t).
\end{equation}
For the spatially uniform coherent phonon considered here, only the zone-center ($\bm{q}=\bm{0}$) term contributes, so $Q^\text{co}_{\bm{q}\nu}(t)$ reduces to the coherent coordinate $Q^\text{co}_{\nu}(t)$ of Eq.~(\ref{eq:cohQ}). Because Eq.~(\ref{selfenergy}) is linear in $Q^\text{co}_{\nu}(t)$, the coherent-phonon contribution to the band energy oscillates in phase with the coherent displacement. The electronic energies are renormalized as $E_{n\bm{k}}^{\mathrm{QP}}=E_{n\bm{k}}+\mathrm{Re}\,\Sigma_{n\bm{k}}$, a shift directly measurable by time- and angle-resolved photoemission spectroscopy \cite{Lv2019AngleresolvedPhotoemissionSpectroscopy, Jensen2026CoherentPhononDrivenBand}. Figure~\ref{cont}(c) shows the coherent-phonon-induced band shifts near $\bm{K}$ for the higher-density trajectory at 1.0~ps and 1.1~ps, the two times marked in Fig.~\ref{cont}(b). These times correspond to opposite sides of the displacement oscillation, so the band shifts occur in opposite directions relative to the plasma-reference bands and reach the tens-of-meV scale.

Thus, the carrier-density dependence of the $A_{1g}$ coherent phonon is reflected in both lattice and electronic observables: the real-space Te displacement, accessible through reflectivity \cite{Fukuda2024CoherentOpticalResponse}, and the coherent-phonon-induced band renormalization, accessible through time-resolved photoemission \cite{Lv2019AngleresolvedPhotoemissionSpectroscopy, Jensen2026CoherentPhononDrivenBand}.
These signatures should be experimentally accessible over the lower part of the density range investigated. In particular, $n_c\sim 10^{14}$ cm$^{-2}$ corresponds to pump fluences comparable to those used in coherent-phonon experiments on monolayer MoTe$_2$ \cite{Sayers2023StrongCouplingCoherent}. At the highest densities, however, heating, nonlinear effects, or structural damage may become relevant, and our results at these densities serve to expose the trend of the plasma-renormalized response.

\section{Conclusion}
We have extended the first-principles equation-of-motion framework for coherent phonons to photoexcited semiconductors by treating the long-lived electron--hole plasma as both a source of residual lattice force and a modifier of the electronic structure. In monolayer MoTe$_2$, this plasma-renormalized framework reproduces the cDFPT equilibrium Te displacement and captures the density-dependent reshaping of the band extrema, including the crossover from a direct gap at $\bm{K}$ to a smaller indirect gap involving $\bm{Q}$. The photoexcited carrier density acts as a microscopic control parameter for the $A_{1g}$ coherent phonon, with increasing $n_c$ enlarging the plasma-shifted equilibrium and strengthening the initial driving force to produce larger oscillations at fixed phase. The resulting density-controlled response is reflected in both the real-space atomic displacement and the coherent-phonon-induced band renormalization, offering complementary signatures in ultrafast reflectivity and time-resolved photoemission.

\begin{acknowledgments} 
The authors acknowledge financial support from the National Natural Science Foundation of China (Grant Nos. 12574250 and 12104039), the Beijing Natural Science Foundation (Grant No. Z260002), and the National Key R\&D Program of China (Grant No. 2022YFA1403400).
\end{acknowledgments}

\section*{Data Availability}
The data that support the findings of this article are openly available~\cite{data}.

\bibliography{CoPh.bib}

\end{document}